\documentclass{ws-p8-50x6-00}

\input epsf
\begin{document}
\title{The $N^*$ Program at Jefferson Lab - Status and Prospects}
\author{Volker D. Burkert}
\address {Jefferson Lab, 12000 Jefferson Avenue, Newport News, VA23606}

\maketitle

\noindent

\abstracts{I discuss recent results on the electroproduction and 
photoproduction of mesons in the region of non-strange baryon resonances. 
Results on the quadrupole transition from the ground state nucleon to 
the Delta show the importance of the nucleon pion cloud. The excitation of 
the ``Roper'' $N^+_{1/2}(1440)$ is being studied in single pion production, 
and the  excitation of the lowest negative parity state $N^-_{1/2}(1535)$ 
is studied in the $p\eta$ and in the $N\pi$ channels.
First results on electroproduction of $p\pi^+\pi^-$ show intriguing resonance
structure which seems difficult to explain using known baryon resonance 
properties. Searches for resonances in $KY^*$ and in $p\omega$ channels 
also reveal resonance-like behavior. I also briefly address a new avenue 
to pursue $N^*$ physics using exclusive deeply virtual Compton scattering, 
recently measured for the first time at JLAB and DESY.} 

\vspace{1cm}

\section{Introduction}

Nucleon physics today is focussed on understanding the details
of the nucleon spin and flavor structures at varying distances, 
and on the systematics of the baryon spectrum. The latter also 
reveals underlying symmetry properties and internal dynamics.   
 
Resonance electroproduction has rich applications in nucleon structure 
studies at intermediate and large distances. Resonances play an 
important role in understanding the spin structure of the 
nucleon\cite{buli,ioffe} at intermediate and large distances. 
More than 80\%  of the helicity-dependent
integrated total photoabsorption cross section difference (GDH integral) 
is a result of the excitation of the $\Delta(1232)$\cite{buli,mami}. 
At $Q^2$= 1GeV$^2$ about 40\% of the strength in the first moment 
$\Gamma_1^P(Q^2) = \int_0^1{g_1(x,Q^2)dx}$ for
the proton is due to contributions of the resonance 
region\cite{devita,burkert_trieste}. 
Conclusions regarding the nucleon spin structure for $Q^2 < 2$GeV$^2$ 
must therefore be regarded with some scepticism if contributions of 
baryon resonances are not taken into account. 

The nucleon's excitation spectrum has been explored mostly with pion beams. 
Many states, predicted in the standard quark model, have not been seen 
in these studies, possibly many of them decouple from 
the $N\pi$ channel\cite{koniuk}.
Electromagnetic interaction and measurement of multi-pion final states,  
may then be the only way to study some of these states. 
Photoproduction of mesons is being used extensively at LEGS, MAMI, GRAAL, 
ELSA and JLAB. Electroproduction adds sensitivity 
due to the fact that the virtual photon carries linear polarization, 
and due to the possibility of varying the photon virtuality. I will 
mostly discuss recent results in electroproduction from JLAB where 
accurate data have now become available.

\begin{figure}[t]
\vspace{65mm} 
\centering{\includegraphics{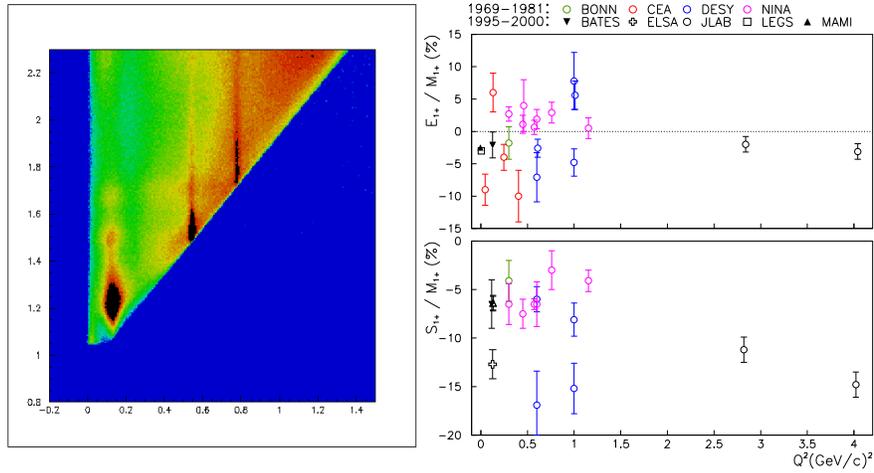}}
\caption{\small Left: Invariant mass $M_{\gamma^* p}$ versus 
 missing mass $M_X$ measured in CLAS. Right: 
$R_{EM}$ and $R_{SM}$ before the year 2001}
\label{epx}
\end{figure}

\begin{figure}[t]
\centering{\includegraphics{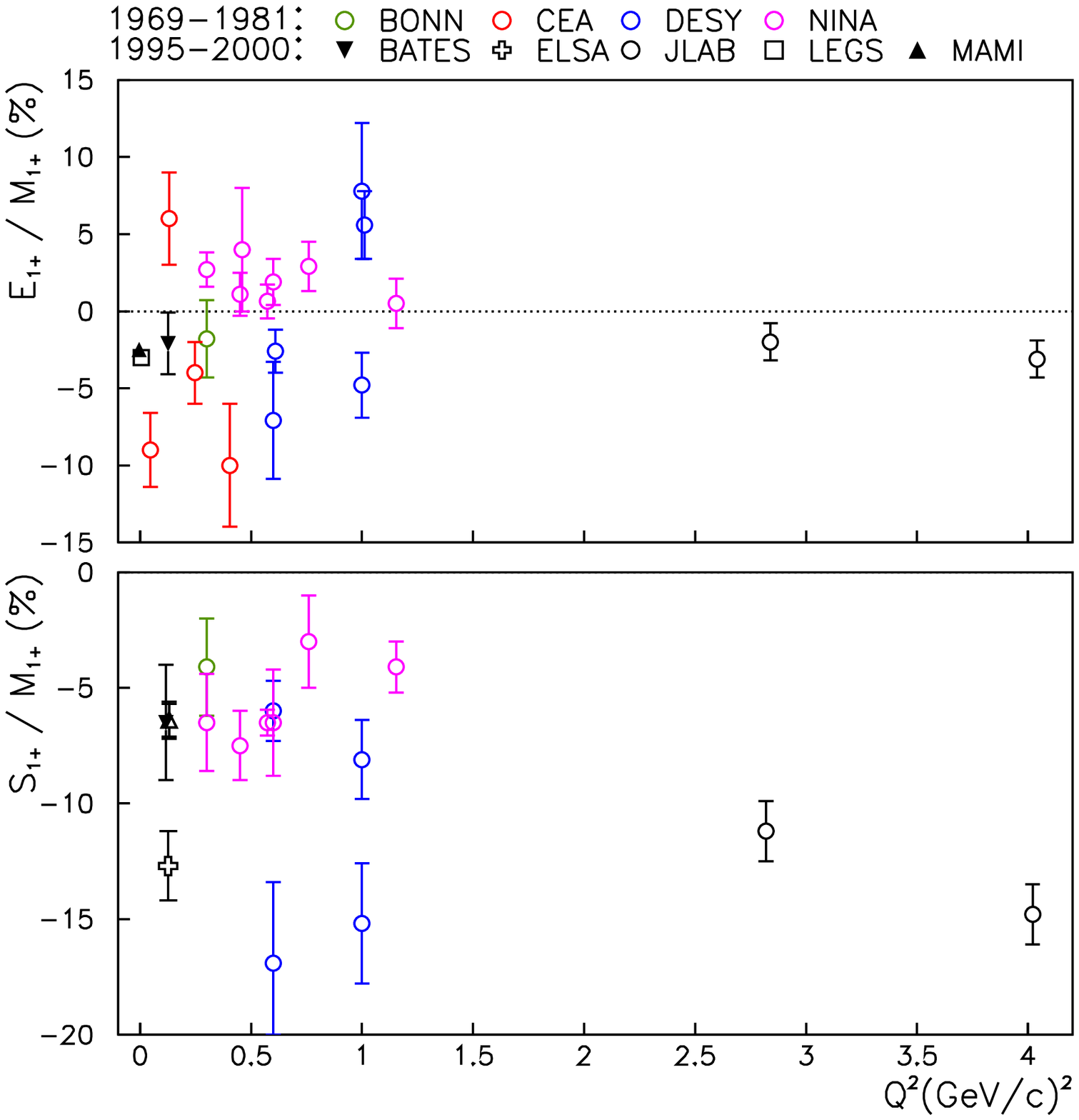}}
\label{delta2001}
\end{figure}

In the past electroexcitation of resonances was not considered a tool of 
baryon spectroscopy but rather a means of measuring transition formfactors 
of some prominent states. CLAS is the first full acceptance instrument with 
sufficient resolution 
to measure exclusive electroproduction of mesons with the goal of studying 
the excitation of nucleon resonances in detail. This feature is illustrated in 
Fig. \ref{epx}, where 
the invariant mass W of the hadronic system is plotted versus the missing
mass of the $ep \rightarrow epX$, where $X$ represents the undetected
system. The $\pi^o$, $\eta$, and 
$\omega$ bands are clearly correlated with enhancements in the invariant 
mass indicating resonances coupling to $p\pi^o$ $p\eta$, and possibly 
$p\omega$ 
channels. Some of the lower mass states are already well known. However, 
their photocouplings and the $Q^2$ evolution of the transition 
form factors may be quite uncertain, or completely unknown.   

\section{Quadrupole Deformations of the Nucleon and the Delta}

An interesting aspects of nucleon structure at low energies 
is a possible quadrupole 
deformation of the nucleon and the $\Delta(1232)$. In some model
interpretation\cite{buchmann} this would be evident in non-zero values of the 
quadrupole transition 
amplitudes $E_{1+}$ and $S_{1+}$ from the nucleon to the $\Delta(1232)$.
In models with $SU(6)$ spherical symmetry, this transition is due to
a magnetic dipole $M_{1+}$ mediated by a simple spin flip 
from the $J= {1 \over 2}$ nucleon ground state to the Delta with 
$J = {3 \over 2}$. 
Non-zero values for $E_{1+}$ would indicate deformation. Dynamically such 
deformations may arise through interaction of the photon with the pion 
cloud\cite{sato,yang} or through the one-gluon 
exchange mechanism\cite{koniuk}. 
At asymptotic momentum transfer, a model-independent prediction of 
helicity conservation 
requires $R_{EM}\equiv E_{1+}/M_{1+} \rightarrow +1$. An interpretation 
of $R_{EM}$ in terms of a quadrupole deformation can therefore only be 
approximately valid at low momentum transfer. 

The data for $R_{EM}$ and $R_{SM}$ published before 2001 
( Fig. \ref{delta2001}) show large systematic 
discrepancies. Taken together no clear trend is visible. However, at the 
photon point recent data from MAMI\cite{beck} and LEGS\cite{legs} 
are consistent with a value of $R_{EM}=-0.0275 \pm 0.005$.

\begin{figure}[t]
\vspace{60mm}
\centering{\includegraphics{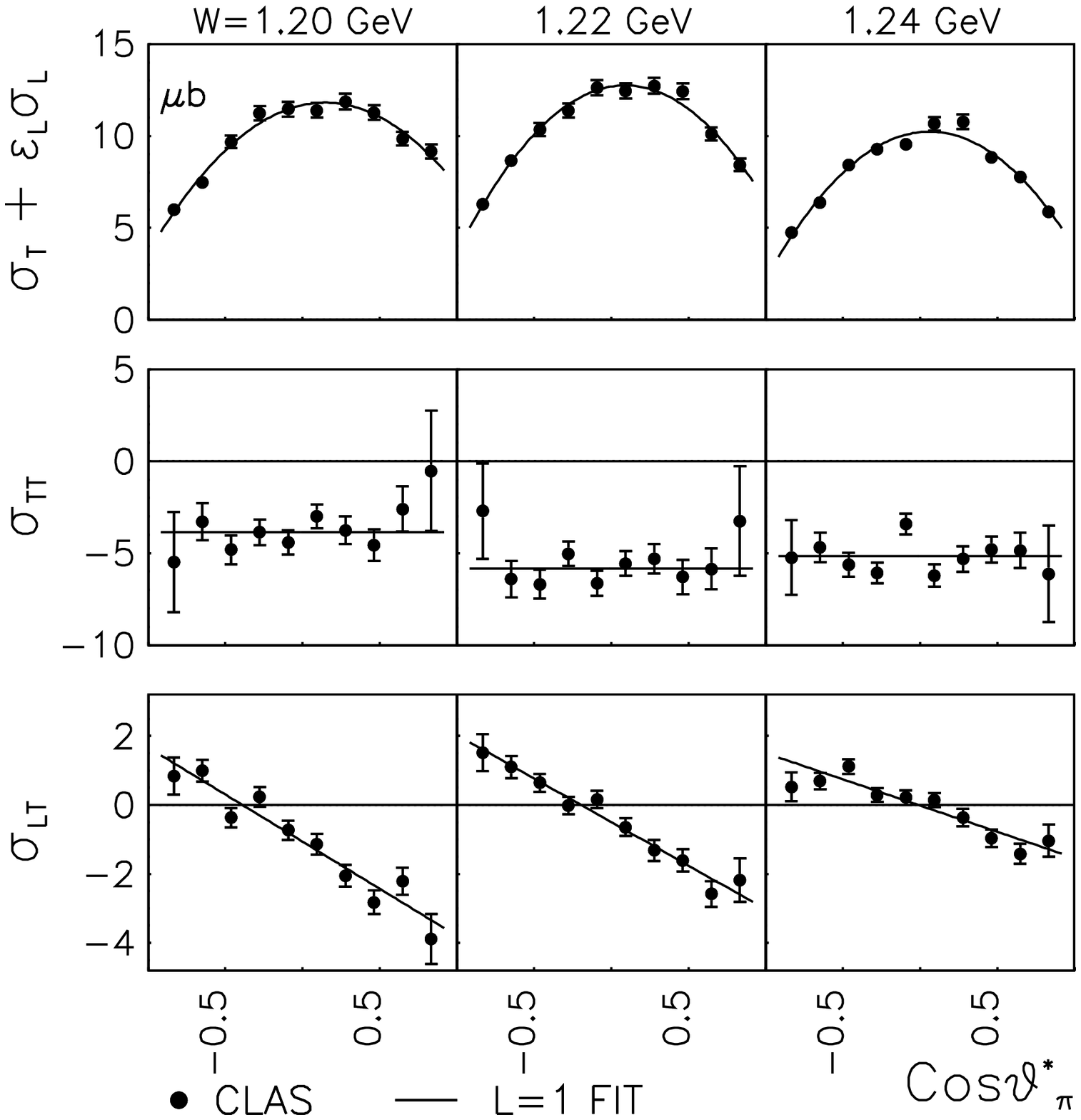}}
\caption{\small Left: Response functions at $Q^2  = 0.9$ GeV$^q$ 
for $p\pi^o$ in the Delta region.
Right: $R_{EM}$ and $R_{SM}$ after 1990, including the latest 
CLAS results covering the range $Q^2$=0.4 - 1.8GeV$^2$.}
\label{deltaresp}
\end{figure}

\begin{figure}[t]
\vspace{-0.5cm}
\centering{\includegraphics{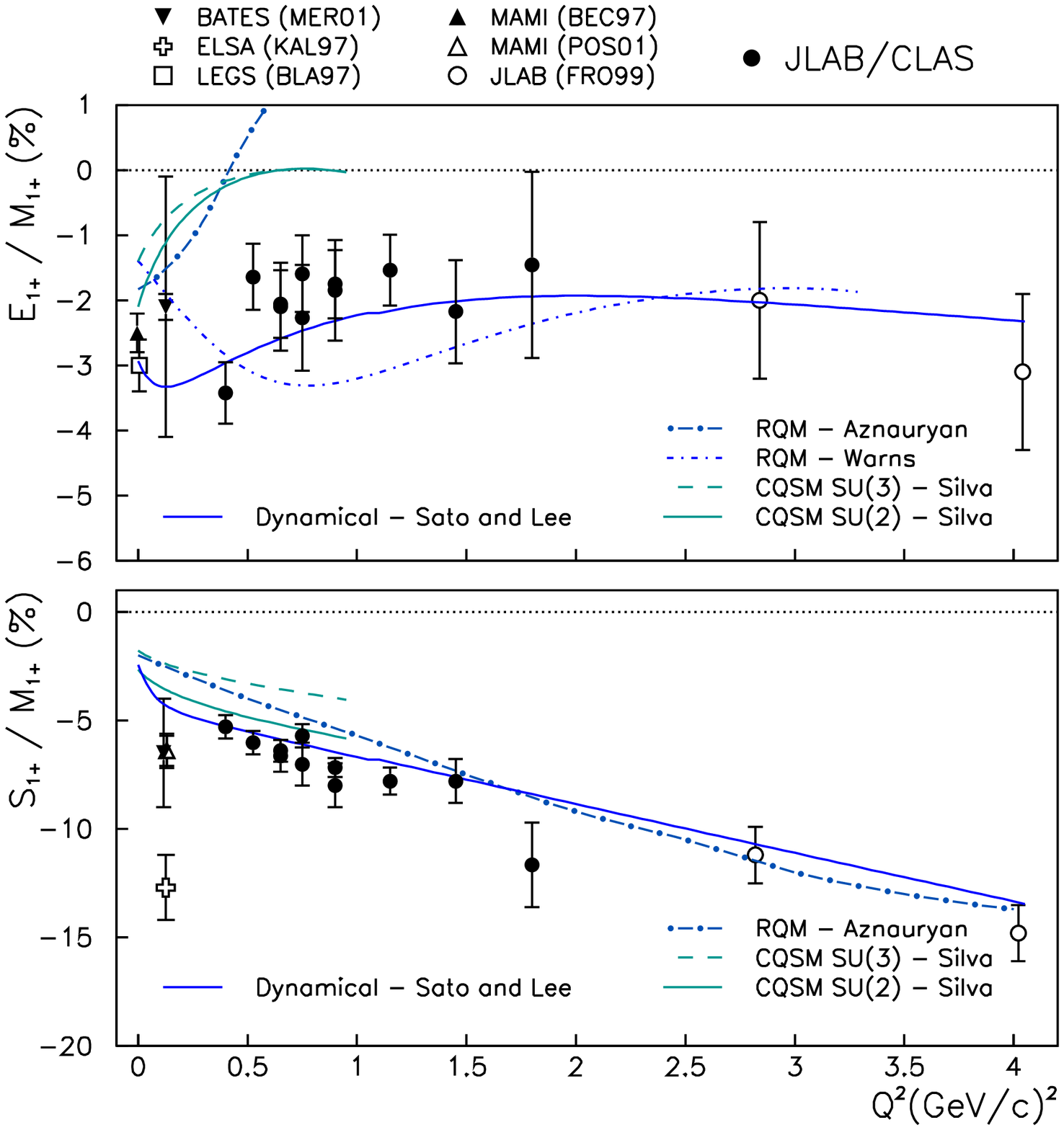}}
\label{remrsm}
\end{figure}

The differential cross section for single pion production is given by
$${d\sigma \over d\Omega_{\pi}} = \sigma_T + \epsilon_L \sigma_L + 
\epsilon \sigma_{TT}\cos{2\phi} + \sqrt{2\epsilon_L(\epsilon+1)}\sigma_{LT}\cos{\phi}
+ h\sqrt{\epsilon_L(1-\epsilon)}\sigma_{LT^{\prime}}\sin{\phi}, \eqno(1)$$
where $\phi$ is the azimuthal angle of the pion, and h=$\pm$1 is the 
helicity of the incident electron. The response functions $\sigma_i$ 
depend on the hadronic invariant mass W, $Q^2$, and polar cms pion angle 
$\theta^*_{\pi}$.   
CLAS allows measurement of the full angular distribution in azimuthal
and in polar angle. The former allows separation of the
$\phi$-dependent and $\phi$-independent response functions $\sigma_i$ 
at fixed $Q^2$, W, and $\cos\theta^*_{\pi}$, while the latter
contains information on the multipolarity of the transition. The multipoles 
can be extracted in some approximation through a fit of  
Legendre polynomials to the angular distribution of the various response 
functions.
Angular distributions of response functions at fixed W values are
shown in Fig. \ref{deltaresp}. Results of the multipole analysis are shown
in Fig. \ref{remrsm}, where data from previous experiments 
published after 1990 are included as well\cite{beck,frolov}. 
$R_{EM}$ remains negative and small throughout the $Q^2$ range. There are 
no indications that leading twist pQCD contributions are important as 
they would result in a rise of $R_{EM} \rightarrow +1$\cite{carlson}. 
$R_{SM}$ behaves quite differently. While it also remains negative, 
its magnitude is strongly rising with $Q^2$. For $Q^2 > 0.35$GeV$^2$ $R_{SM}$
follows approximately a straight line that may be parametrized as:
 $R_{SM} = -(0.04 + 0.028\times Q^2$(GeV$^2$)). 
The comparison with 
microscopic models, from relativized quark models\cite{warns,aznaury},
chiral quark soliton model\cite{silva}, and dynamical 
models\cite{sato,yang} show that simultaneous description of both $R_{EM}$
and $R_{SM}$ is achieved by dynamical models that include the 
nucleon pion cloud, explicitely. This supports the claim that most of the 
quadrupole strength is due to meson effects which 
are not included in other models. 

\begin{figure}[t]
\vspace{75mm}
\centering{\includegraphics{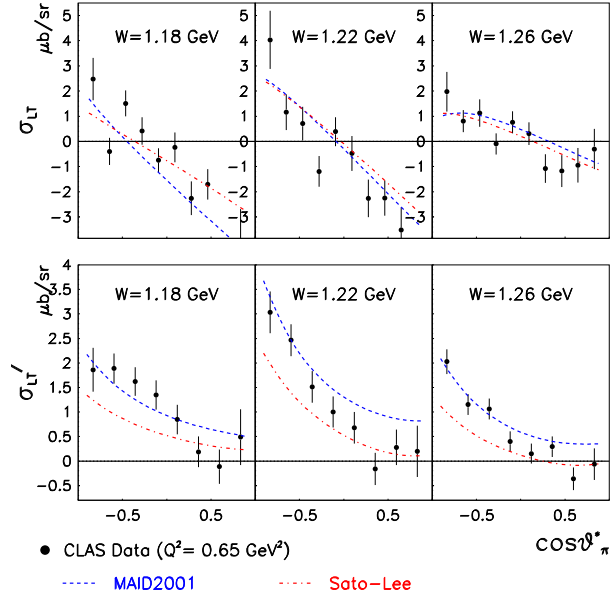}}
\hspace{5cm} 
\caption{\small Top panel shows sesponse functions $\sigma_{LT}$ for 
$\pi^0$ production in the $Delta(1232)$ region. At the peak of 
the $\Delta$ there is virtually no model-dependency. The bottom panel 
shows preliminary $\sigma_{LT^{\prime}}$ data from CLAS. 
These data shows strong model sensitivity at the peak of the $\Delta$.  
The curves are model predictions of Sato-Lee, and MAID2000.}
\label{clasrltp}
\end{figure}

The new CLAS data establish a new level of accuracy. However, 
improvements in statistical precision and the coverage of a larger 
$Q^2$ range are expected for the near future, and they 
must be complemented by a reduction of model dependencies in the analysis. 
Model dependencies are largely due to our poor knowledge of the 
non-resonant contributions, which become increasingly important at higher $Q^2$. The 
$\sigma_{LT^{\prime}}$ response function, a longitudinal/transverse 
interferences term is especially sensitive to 
non-resonant contributions if a strong resonance is present. 
$\sigma_{LT^{\prime}}$ can be measured using a polarized electron beam 
in out-of-plane kinematics for the pion. Preliminary 
data on $\sigma_{LT^{\prime}}$ from CLAS are shown in 
Fig. \ref{clasrltp} in comparison with dynamical models, clearly showing
the model sensitivity to non-resonant contributions. 
Both models predict nearly the same unpolarized 
cross sections, however they differ in their handling of non-resonant 
contributions.

\subsection{The $\gamma N \Delta(1232)$ transition in lattice QCD}

Ultimately, we need to come to a QCD description of these important 
nucleon structure quantities. Currently, there is only one calculation 
in quenched
 QCD\cite{leinweber} giving  $R_{EM} = 0.03 \pm 0.08$, which, due to the 
large uncertainty, has little bearing on our current understanding of nucleon
structure. This calculation was made nearly a decade ago. 
Improvements in computer performance and improved QCD actions 
and lattices should allow a reduction of the error to a level
where QCD should provide significant input. Such calculations seem feasible
now and are are part of the program of the Lattice Hadron Physics 
Collaboration\cite{lattice-2}. They are urgently needed to link these 
fundamental quantities directly to QCD.

\section{$N^*$'s in the Second Resonance Region}

Three states, the elusive ``Roper'' $N^{\prime}_{1/2^+}(1440)$, 
and two strong negative parity states, $N^{*}_{3/2^-}(1520)$, 
and $N^{*}_{1/2^-}(1535)$ make up the 
second enhancement seen in inclusive electron scattering. All of these 
states are of special interest to obtain a better understanding of 
nucleon structure and strong QCD.

\begin{figure}[t]
\vspace{65mm}
\centering{\includegraphics{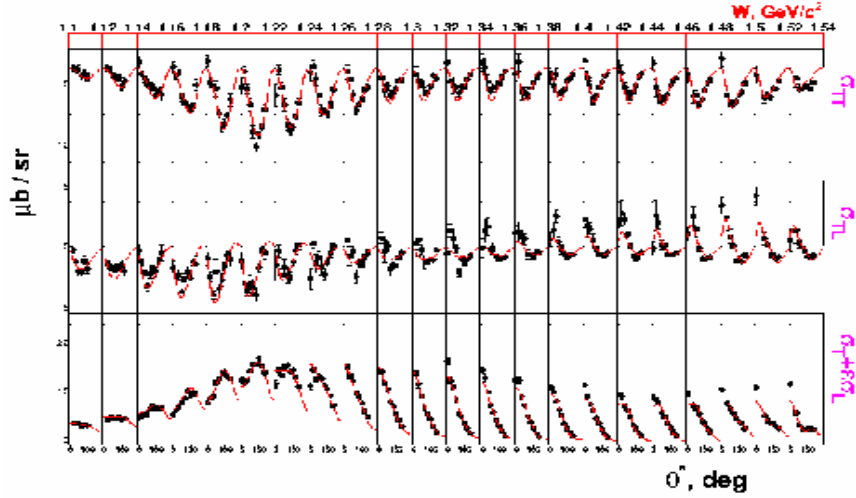}}
\hspace{5cm} 
\caption{\small  Response functions for $\gamma^* p \rightarrow n\pi^+$. 
The data cover the $\Delta(1232)$ and the 2nd resonance regions. Angular 
distributions are show for each bin in W. The 
curve represent a fit of a unitary isobar model.}
\label{npipl}
\end{figure}

\subsection{The Roper resonance - still a mystery}

The Roper resonance has been a focus of attention for the last decade, largely due
to the inability of the standard constituent quark model to describe basic
features such as the mass, photocoupling, and $Q^2$ evolution of the transition 
form factors. 
This has led to alternate approaches where the state is assumed to have 
a strong gluonic component\cite{libuli}, a small quark core with a large 
meson cloud \cite{cano}, or a hadronic molecule of a nucleon and a 
hypothetical $\sigma$ 
meson $|N\sigma>$\cite{krewald}. Recent LQCD calculations\cite{lattice-1,lattice-2,lattice-3} 
show no sign of a 3-quark state with the quantum numbers of the nucleon 
in the mass range of the Roper state.  

\begin{figure}[t]
\vspace{60mm}
\centering{\includegraphics{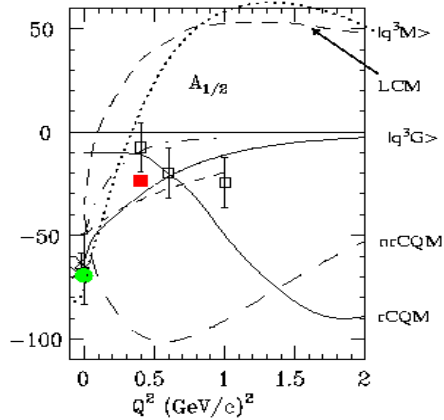}}
\caption{\small  Transverse helicity amplitude $A_{1/2}(Q^2)$ for the Roper 
resonance. The full squared red symbol is a preliminary point from CLAS 
(see text). Models include a non-relativistic constituent 
quark model (nrCQM), 
relativized model (rCQM), hybrid baryon 
model ($|q^3G>$), a model in light cone kinematics (LCM), 
and a model containing a small quark core and meson cloud ($|q^3M>$).}
\label{roper}
\end{figure}

Experimentally, the Roper as a isospin 1/2 state couples more strongly
to the n$\pi^+$ channel than to the p$\pi^o$ channel. Lack of data in
that channel and lack of polarization data has hampered progress in the past.
Fortunately, this sitation is changing significantly with the new data
from CLAS. For the first time complete angular distributions
have been measured for the n$\pi^+$ final state. Preliminary separated 
response functions obtained with CLAS are shown in Fig. \ref{npipl}.
These data, together with the $p\pi^o$  response functions, as well as 
the spin polarized $\sigma_{LT^{\prime}}$ response function for both channels 
have been fitted to a unitary isobar model\cite{janr}. 
The preliminary results are shown in Fig. \ref{roper} together with the sparse 
data from previous analyses. The CLAS data confirm the rapid drop of $|A_{1/2}|$
with $Q^2$. From all models only the hybrid model\cite{libuli} is 
somewhat consistent with the measured $Q^2$ evolution, 
although
much improved data are needed for more definite tests in a 
large $Q^2$ range. An important question is if the $A_{1/2}(Q^2)$ amplitude
changes sign, or remains negative. Also, all models which assume a 3-quark 
structure at short distances predict a very slow $Q^2$ dependence of 
$A_{1/2}$, while the gluonic model predicts a fast fall-off with $Q^2$. 
The range of model 
predictions for the $Q^2$ evolution illustrates dramatically the 
sensitivity of electroproduction to the internal structure of this 
state. Model builders should therefore implement electromagnetic couplings 
and their $Q^2$ evolution into their models, as experiments can provide
most stringent tests.

\subsection{The first negative parity state $N^*_{1/2^-}(1535)$}

Another state of interest in the 2nd resonance region is the 
$N^*_{1/2^-}(1535)$. This state was found to have an unusually 
hard transition formfactor, i.e. the $Q^2$ evolution shows a slow
fall-off. This state is often studied in the $p\eta$ channel and
shows a strong s-wave resonance near the $\eta$-threshold with very little
non-resonant background. Older data show some discrepancies as to the 
total width and photocoupling amplitude. In particular, analysis of 
$N\pi$ photoproduction data\cite{pdg} disagree with the analyses of
the $\eta$ photoproduction data by a wide margin.

\begin{figure}[t]
\vspace{60mm}
\centering{\includegraphics{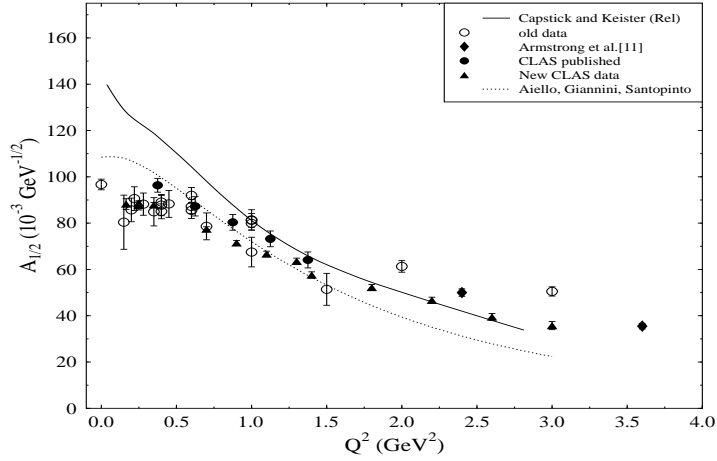}}
\hspace{5cm} 
\caption{\small  Transverse helicity amplitude $A_{1/2}(Q^2)$ for the 
first negative parity state $N^*_{1/2^-}(1535)$.} 
\label{s11}
\end{figure}

Data from CLAS\cite{thompson}, together with 
data from an earlier JLab experiment\cite{armstrong} give now a 
consistent picture of 
the $Q^2$ evolution, confirming the hard formfactor behavior with 
much improved data quality, as shown in Fig. \ref{s11}.  Analysis of
the n$\pi^+$ and p$\pi^o$ data at $Q^2$= 0.4GeV$^2$ gives a value for 
$A_{1/2}$ consistent with the analysis of the p$\eta$ data at the same $Q^2$.
Therefore, there is no discrepancy in electroproduction between the 
$p\eta$ and the $N\pi$ data analyses. 

There is also some good news at the theory level. The hard form factor 
has been difficult to understand in any model. Recent work  
within a constituent quark model using a hypercentral 
potential\cite{santopinto} has made significant progress in reproducing the  
$A_{1/2}$ amplitude for the $N^*_{1/2^-}(1535)$. The hard form factor 
is also in contrast to models that interpret this state as a $|\bar K\Sigma>$
hadronic molecule\cite{weise}. Although no calculations 
exist for such models, the extreme ``hardness'' of the formfactor and the 
large cross section appear counter intuitive to an 
interpretation of this state as a bound hadronic system.
Lattice QCD calculations also show very clear 3-quark strength  
for the state\cite{lattice-1,lattice-2,lattice-3}, making it an 
unlikely candidate for such a hadronic system.

\section{Higher Mass States and Missing Resonances}

Approximate $SU(6)\otimes O(3)$ symmetry of the symmetric constituent 
quark model leads to relationships between the various states. In the 
single-quark transition model (SQTM) a single quark participates in the 
interaction. The model predicts transition amplitudes for a large number of
states based on only a few measured amplitudes.  
Electroproduction data for states with masses near 1.7 GeV, many assigned to
the $[70,1^-]$ supermultiplet, are needed for 
significant tests of this model. Unfortunately, these states
decouple largely from the $N\pi$ channel but 
couple dominantly to the $N\pi\pi$ channel for which 
no photo- or electroproduction data exist. 
Moreover, many of the so-called ``missing'' states are predicted to 
couple strongly to the $N\pi\pi$ channels\cite{capstick2}. 
Study of $\gamma^*p \rightarrow p\pi^+\pi^-$ 
as well as the other charge channels are therefore important also in the 
search for undiscovered baryon states. This is of particular importance 
for our understanding of baryons structure as many of the ``missing'' states 
are not predicted in alternative models that are not
based on approximate $SU(6)\otimes O(3)$ symmetry\cite{kirchbach}. 

\begin{figure}
\vspace{100mm}
\centering{\includegraphics{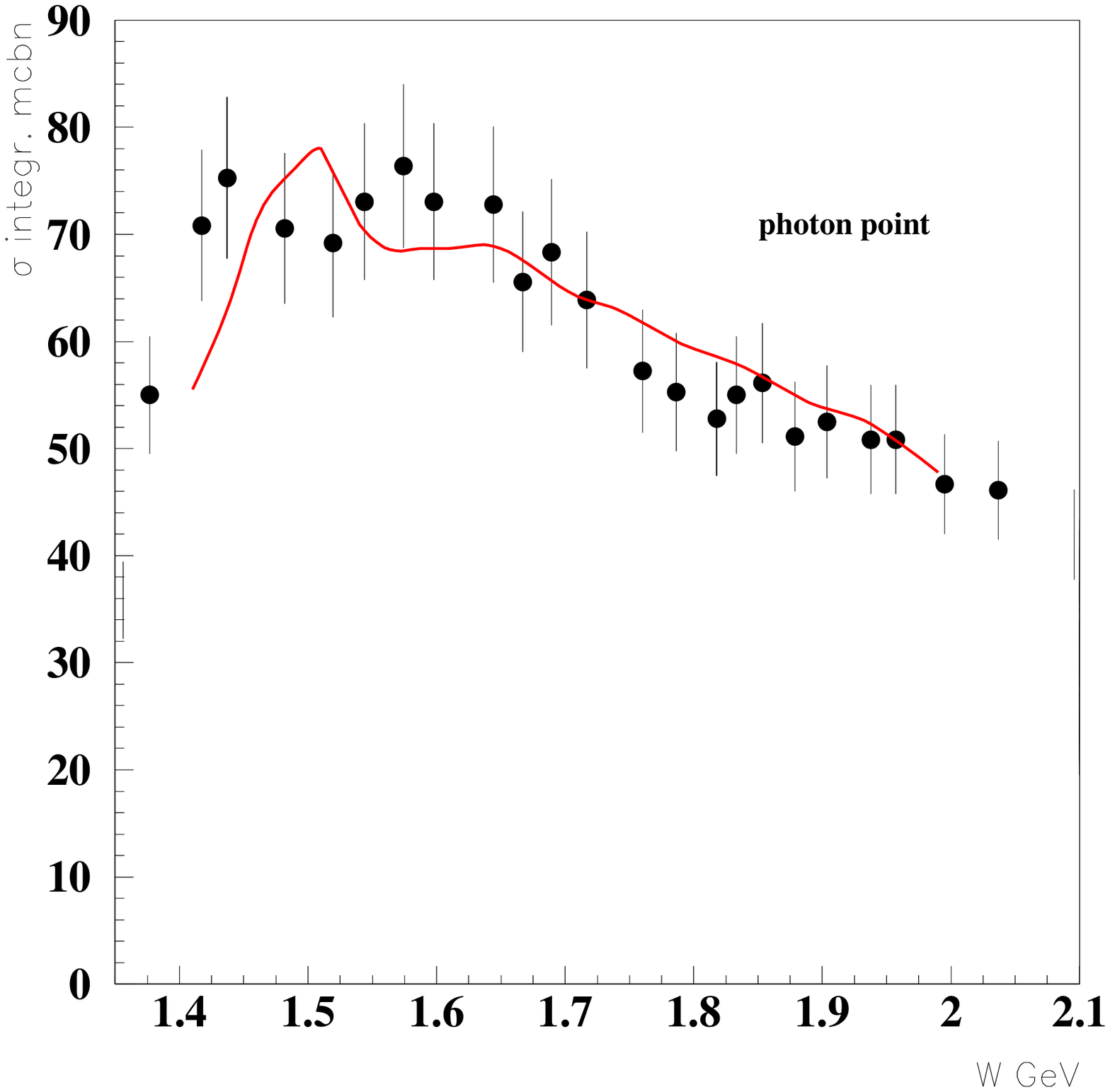}}
\centering{\includegraphics{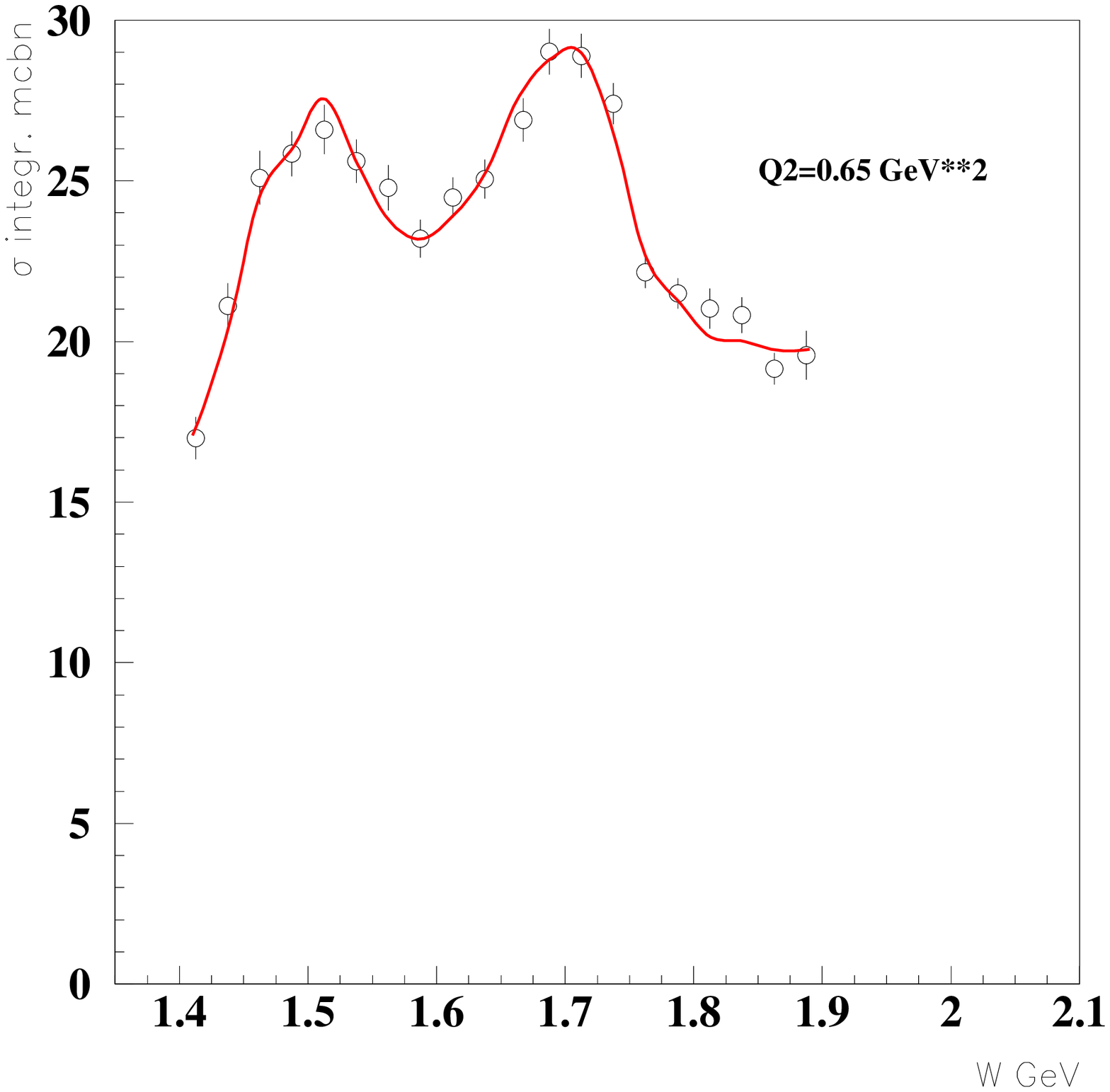}}
\centering{\includegraphics{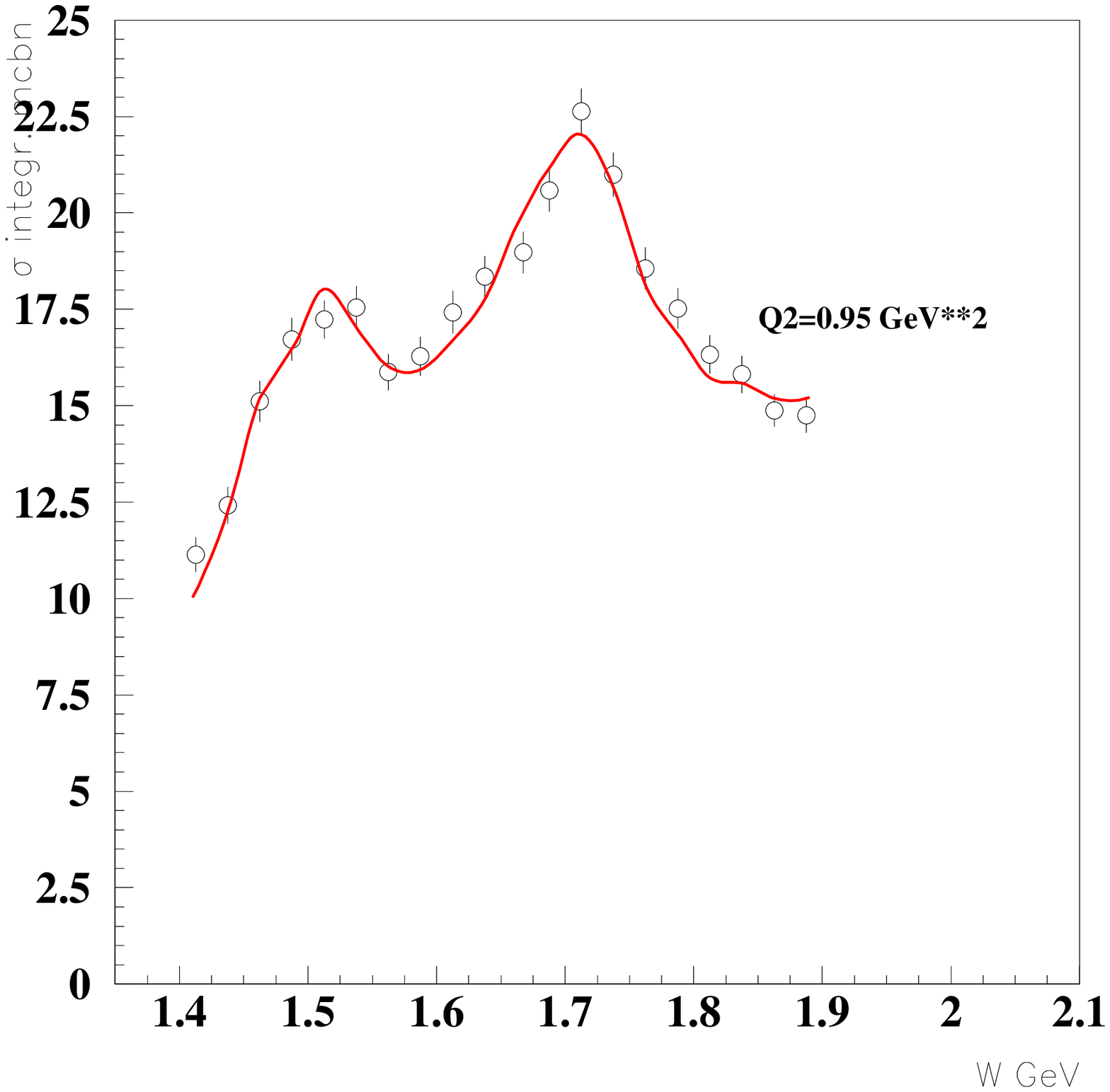}}
\centering{\includegraphics{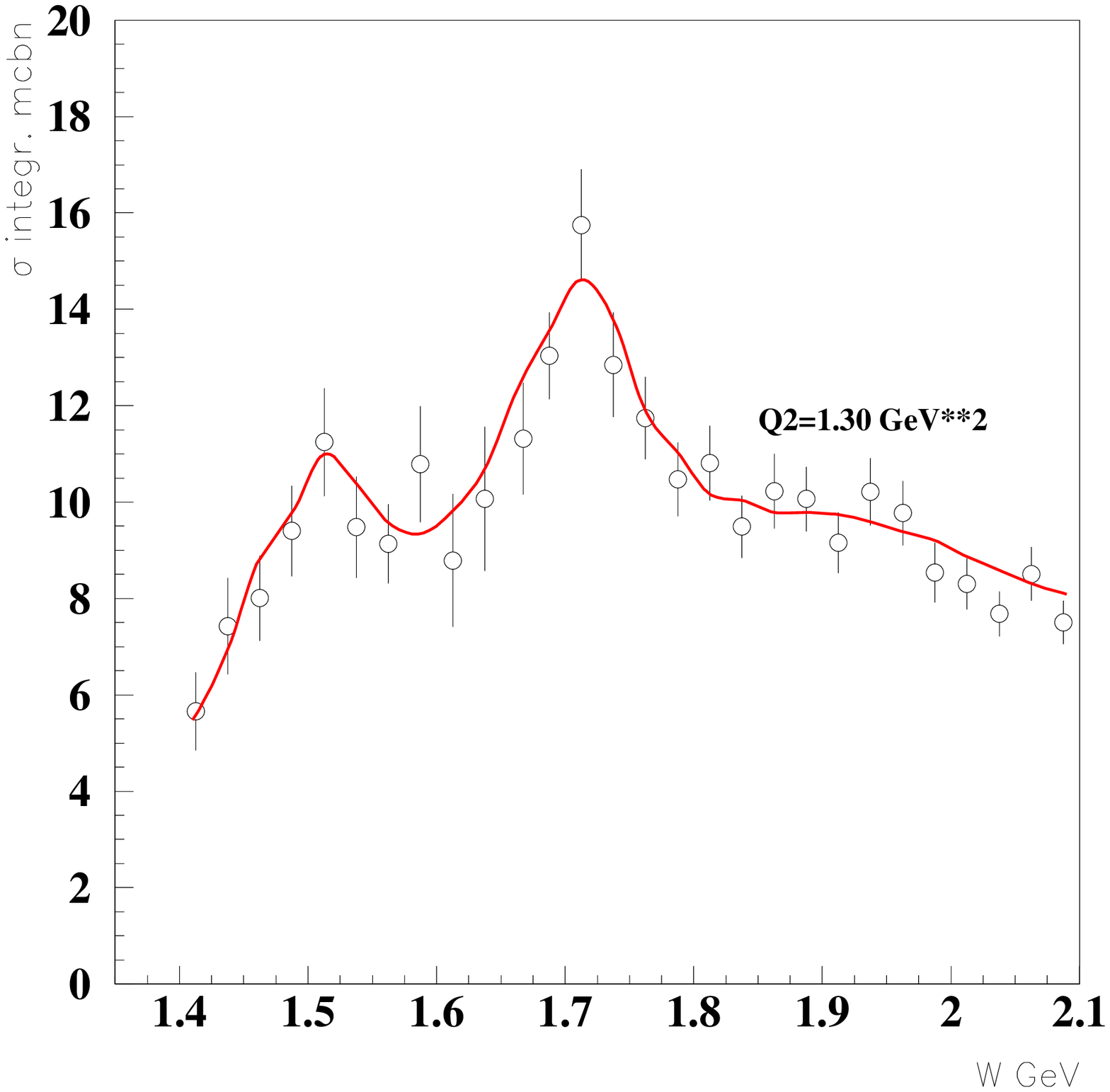}}

\caption{\small Total photoabsorption cross section for 
$\gamma^* p \rightarrow p\pi^+\pi^-$. Photoproduction data from DESY - top
left panel. The other panels show CLAS electroproduction data 
at $Q^2=0.65~,0.95,~1.3 GeV^2$ }
\label{ppippim}
\end{figure}

\subsection{A new $N^*_{3/2^+}(1720)$ quark model state?}

New CLAS total cross section electroproduction data are shown in 
Fig. \ref{ppippim} in comparison with photoproduction 
data from DESY.
The most striking feature is the strong resonance peak near W=1.72 GeV
seen for the first time in the electroproduction of the $p\pi^+\pi^-$ channel. 
This peak is absent in the DESY photoproduction data. The CLAS 
data\cite{ripani} also contain the complete hadronic 
angular distributions and $p\pi^+$ and $\pi^+\pi^-$ 
mass distributions over the full W range. They have been analyzed and the
resonance near 1.72GeV was found to be best described by a $N^*_{3/2^+}(1720)$
state. While there exists a state with such quantum numbers in this mass range,
its hadronic properties were previously found to be very different from
the resonance seen in the CLAS data. 
The PDG gives for the known state a 
$N\rho$ coupling of $\Gamma_{N\rho}/\Gamma_{tot}(PDG) \approx 0.85$ 
while the resonance in the CLAS data has a very small coupling to that channel 
$\Gamma_{N\rho}/\Gamma_{tot}(CLAS) \approx 0.17$. Also, other parameters
such as $\Gamma_{\Delta\pi}/\Gamma_{tot} = 0.62 \pm 0.12$ are remarkably 
different from previous studies of the PDG state.
The question arises if the state could be one of the ``missing'' states. 
Capstick and Roberts\cite{capstick2} predict a second state with these quantum 
numbers at a mass approximately 1.85GeV. There are also predictions of a hybrid baryon
state with these quantum numbers at about the same mass\cite{page}, although 
the rather hard form factor as seen in Fig. \ref{ppippim} disfavors such an 
interpretation\cite{libuli}.
As mass predictions in these models are uncertain to at least
$\pm$100MeV, interpretation of this state as a ``missing'' $N^*_{3/2^+}$  
is a definite possibility. 
Independent of possible interpretations, the hadronic properties of the state 
seen in the CLAS data 
are incompatible with properties of the known state with same 
quantum numbers as listed in Review of Particle Properties\cite{pdg}. 
There is no
a priori reason why the two states should be identical as the PDG state has been 
studied in $\pi N$ scattering, and must therefore couple to $\pi N$, 
while the state seen in the CLAS data only requires coupling to virtual photons 
and the $N\pi\pi$ channel. More definite conclusions may be drawn when 
the analysis of single pion electroproduction data in this mass range 
have been completed.

\begin{figure}
\vspace{50mm}
\centering{\includegraphics{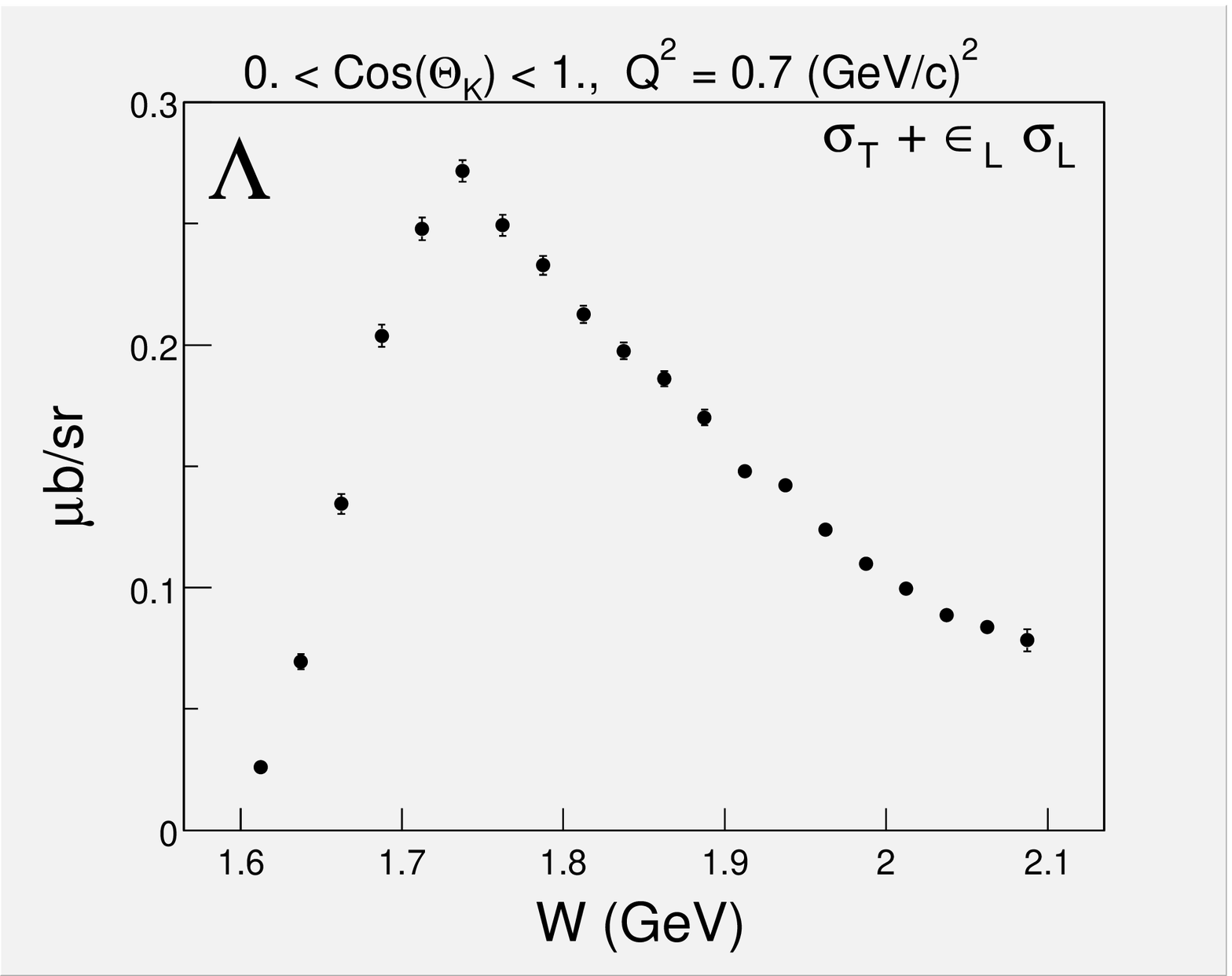}}
\centering{\includegraphics{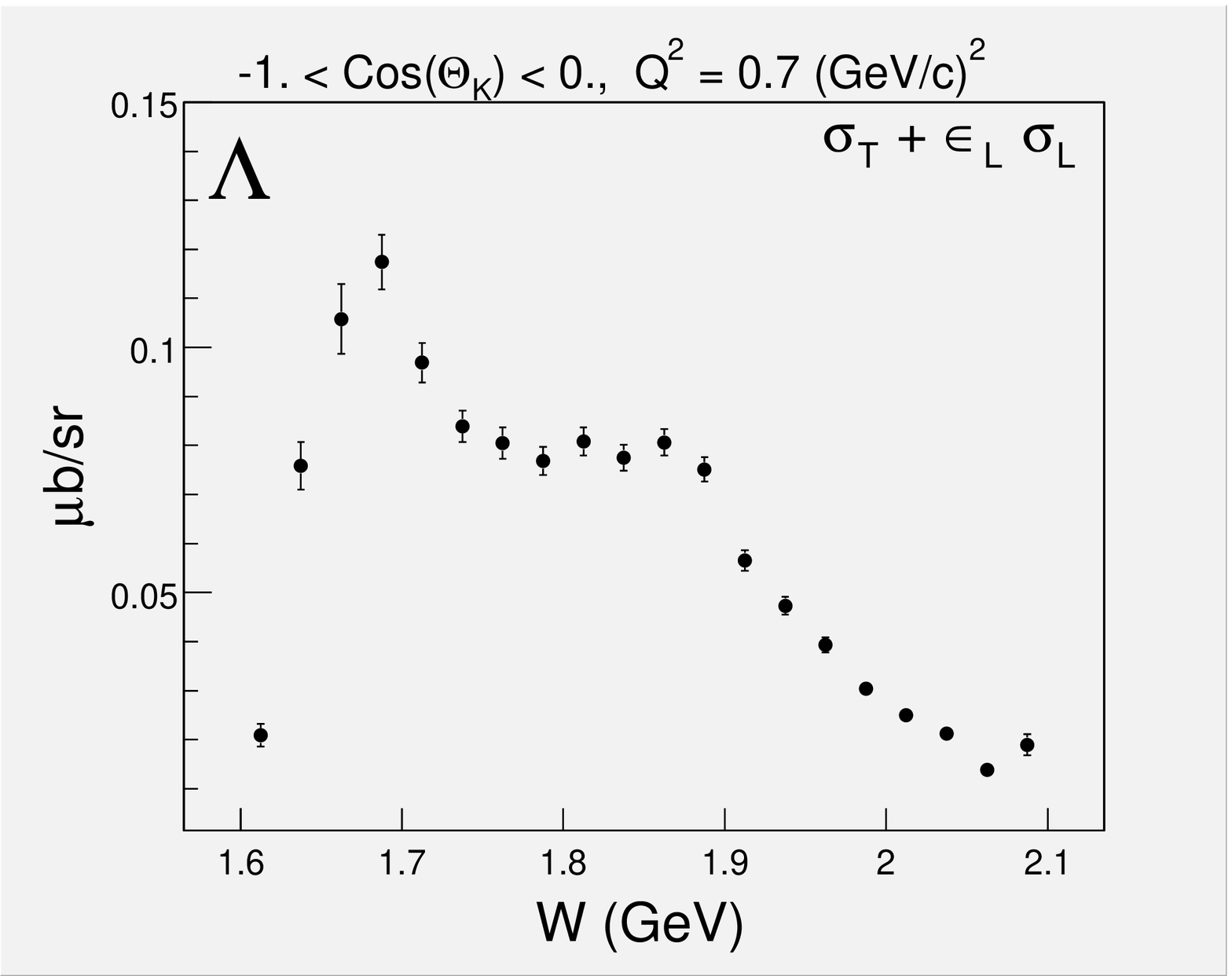}}
\caption{\small Total photoabsorption cross section measured with CLAS for 
$\gamma^* p \rightarrow K^+\Lambda$. The left panel is integrated over the full
forward hemisphere in the $K^+$ angular distribution in the $K^+\Lambda$ cms. 
The right panel is integrated over the backward hemisphere.}
\label{klambda}
\end{figure}

\subsection{Hard nucleon spectroscopy}

The isobar analysis of the $p\pi^+\pi^-$ data in Fig. \ref{ppippim} 
shows that the cross section ratio 
R= resonance/background at W = 1.72 GeV is strongly rising with $Q^2$. 
Electron scattering, especially at relatively 
high photon virtuality, $Q^2$, can therefore provide much increased 
sensitivity in 
the study of higher mass resonances. Quantitatively, this can be 
understood within a non-relativistic dynamical quark model 
\cite{cko,koniuk}. Photocoupling amplitudes for these states 
contain polynomials of powers of the photon 3-momentum vector 
$|\vec q|$. 
For virtual photons the 3-momentum $|\vec q|$ for the transition to a 
given resonances 
increases with $Q^2$ leading to the enhancement. In the case of the 
$N^*_{3/2^+}$ the non-relativistic 
quark model predicts (modulo a formfactor which is common to all states) 
$A_{1/2} = C(|\vec q| + |\vec q|^3/3\alpha^2)$, where $\alpha$ 
is the harmonic oscillator constant of the model.

To the degree that non-relativistic kinematics can be applied, 
spectroscopy of higher mass states with virtual photons at higher 
photon virtualities (``hard spectroscopy'') has a distinct advantage 
over the real photon case: as the power n in $|\vec q|^n$ depends 
on the specific $SU(6)\otimes O(3)$ multiplet a state is 
associated with, it allows enhancing the excitation strength of some 
states over others by selecting specific $Q^2$ ranges.

\subsection{Nucleon states in $K\Lambda$ production}

Strangeness channels have recently been examined in photoproduction
as a possible source of information on new baryon states,
and candidate states have been discussed\cite{saphir,d'angelo}.
The analysis of the 
$K\Lambda$ channel is somewhat complicated by the large t-channel 
exchange contribution producing a peak at forward angles.
Pleliminary electroproduction data have become available from CLAS\cite{niculescu}. 
To increase sensitivity to s-channel processes the data have been 
divided into a set for the forward hemisphere
and a set for the backward hemisphere. While the cross section in the 
forward hemispere shows a t-channel like smooth behavior, clear resonance 
structures emerge in the invariant hadronic mass 
for the backward hemisphere (right panel in Fig. \ref{klambda}).
The lower mass peak near 1.7 GeV is likely due to known resonances; e.g. the 
$P_{11}(1710)$,  while the peak 
near 1.85GeV has no correspondence in the PDG summaries, and 
maybe associated with the candidate state seen first in the SAPHIR 
detector\cite{saphir} at 1.93 GeV although the mass seems to be 
considerable lower. 
A more complete analysis of the angular distribution and of the 
energy-dependence is needed for more definite conclusions.

\subsection{Study of $\gamma p \rightarrow p\eta$}

\begin{figure}
\vspace{70mm}
\centering{\includegraphics{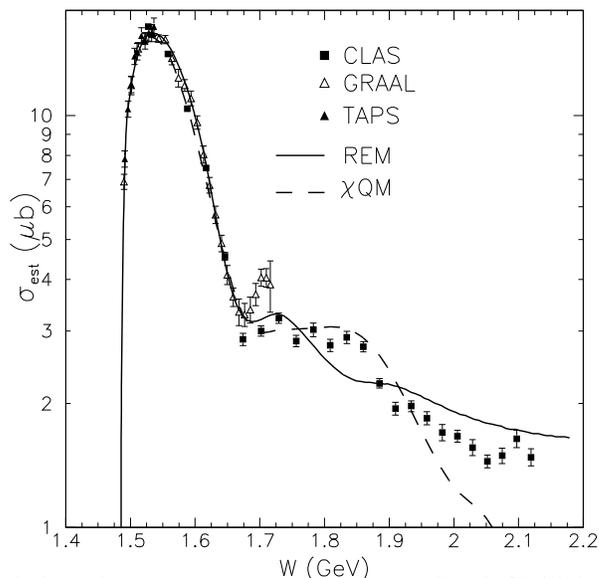}}
\caption{\small Total photoabsorption cross section measured with CLAS for 
$\gamma p \rightarrow \eta p$. The curves represent calculations within a Chiral Quark Model
 and  Eta-MAID (REM). }
\label{eta_photo}
\end{figure}

New $p\eta$ photoproduction data from CLAS\cite{asu} cover now the mass range from threshold up 
to W = 2.1 GeV. Nearly complete angular distributions have been measured for all W bins. 
The total cross section data are shown in Fig. \ref{eta_photo}. While there is good agreement
with previous data from MAMI and GRAAL for W $<$ 1.675 GeV, there is disagreement
with the GRAAL data near 1.7 GeV. The sharp rise near the end point of the GRAAL data 
is not seen in the CLAS data. The broad structure in the mass range near 1.8 GeV
indicates possible resonance excitations. A more complete analysis of the angular
distributions and the energy-dependences is needed for more definite conclusions
regarding higher mass resonance contribution in this channel.

\subsection{Resonances in Virtual Compton Scattering}

Virtual Compton Scattering, i.e. $\gamma^* p \rightarrow p\gamma$ is yet another tool
in the study of excited baryon states. This process has recently been measured in
experiment 
E93-050 in JLAB Hall A\cite{fonvieille} at backward photon angles. The 
excitation spectrum exhibits clear resonance excitations in the mass regions
of known states, the $\Delta(1232)$, $N^*(1520)$, and $N^*(1680)$. The main
advantage of this purely electromagnetic process is the absence of final state 
interaction, which in meson production complicates the interpretation of the 
data. A disadvantage is the low rate which makes it difficult to collect 
sufficient statistics for a full partial wave analysis. 

\begin{figure}
\vspace{90mm}
\centering{\includegraphics{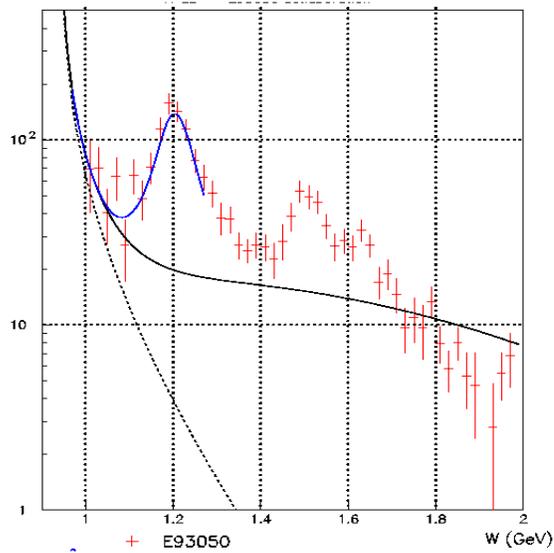}}
\caption{\small Differential cross section for virtual Compton scattering at 
$Q^2=1$ GeV$^2$. The final state photon is in the backward direction relative
to the virtual photon .}
\label{vcs}
\end{figure}

\section{Baryon spectroscopy at short distances}

Inelastic exclusive virtual Compton scattering in the deep inelastic regime (DVSC) 
opens up a new avenue of resonance studies at the elementary quark level.
The process of interest is $\gamma^* p \rightarrow \gamma N^* (\Delta^*)$ 
where the virtual photon is deeply virtual ( $Q^2 >> 1$GeV$^2$).
The photon couples to an elementary quark with momentum 
fraction $x+\xi$, which is re-absobed into the baryonic system with a 
momentum fraction $x-\xi$, after having emitted a high energy photon. 
The recoil baryon system may be a ground state proton or an excited state. 
The elastic exclusive DVCS process has recently been measured with CLAS 
at JLab\cite{stepanya} 
and at DESY\cite{hermes} in polarized electron proton scattering, 
and the results are consistent with predictions from perturbative QCD and the
twist expansion for the process computed at the quark-gluon level. 
The theory is under control for small momentum transfer to the final state 
baryon. For the inelastic process, where a $N^*$ or 
$\Delta$ resonance is excited, the process can be used to study 
resonance transitions at the elementary quark level. Varying parameter 
$\xi$ and the momentum transfer $t$ to the baryonic system, allows probing the 
correlation functions or generalized parton 
distributions (GPDs).

\begin{figure}
\vspace{60mm}
\centering{\includegraphics{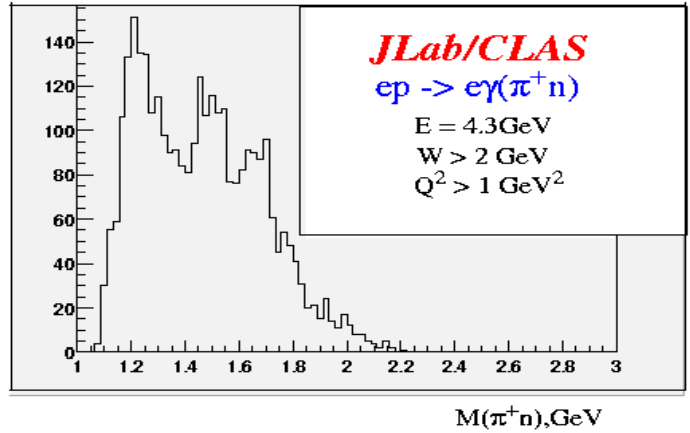}}
\caption{\small Inelastic deeply virtual Compton scattering measured 
in CLAS. The recoiling $(n\pi^+)$ system shows the excitation of
several resonances, the $\Delta(1232)$, $N^*(1520)$, and $N^*(1680)$. }
\label{deltadvcs}
\end{figure}

That this process is indeed present at a measureable level is demonstrated 
with the preliminary data from CLAS shown in Fig. \ref{deltadvcs}. 
The reaction is measured at invariant masses $W > 2$ GeV of the hadronic system. 
The recoiling baryonic system clearly  
shows the excitation of resonances, the $\Delta(1232)$, $N^*(1520)$, 
and $N^*(1680)$. While these are well known states which are also excited in 
the usual s-channel processes, the DVCS process has the advantage 
that it decouples the photon virtuality $Q^2$ from the 4-momentum 
transfer $t$ to the baryon system. $Q^2$ must be chosen sufficiently high 
such that the virtual photon couples to an elementary quark, while the 
momentum transfer to the nucleon system can be varied independently. 
In this way, a theoretical framework employing 
perturbative methods can be used to probe the ``soft'' $NN^*$ transition,
allowing to map out internal parton correlations for this transition.

\section{Conclusions} 

Electroexcitation of nucleon resonances has evoled to an 
effective tool in studying nucleon structure in the regime of 
strong QCD and confinement. The new data from JLab in the 
$\Delta_{3/2^+}(1232)$ and $N^*_{1/2^-}(1535)$ regions give a
consistent picture of the $Q^2$ evolution of the transition 
form factors. Large data sets in different channels including
polarization observables will vastly improve the analysis of
states such as the ``Roper'' $N^{\prime}_{1/2^+}(1440)$, and 
many other higher mass states. A strong resonance signal near 1.72 GeV, 
seen with CLAS in the $p\pi^+\pi^-$ channel, exhibits hadronic 
properties which are incompatible with those of known states 
in this mass region, and may require introduction of 
a ``missing'' baryon state. 

While s-channel resonance excitation  
remains the backbone of the $N^*$ program for years to come, 
inelastic deeply virtual Compton scattering is a promising new 
avenue in 
resonance physics at the elementary parton level, allowing the study 
of parton correlations within a well defined theoretical framework.



\begin{thebibliography}{200} 
\bibitem{buli} V.D. Burkert, and Zh. Li, Phys.Rev.D47:46-50,1993 
\bibitem{ioffe}V.D. Burkert and B.L. Ioffe, Phys.Lett.B296: 223-226,1992; \\
J.Exp.Theor.Phys.78:619-622,1994 
\bibitem{mami}J. Ahrens et al., Phys.Rev.Lett.87:022003,2001 
\bibitem{devita} R. De Vita, Talk at Baryons2002, Jefferson Lab, 
March 3-8, 2002
\bibitem{burkert_trieste} V.D. Burkert, Nucl.Phys.A699:261-269,2002 
\bibitem{koniuk} R. Koniuk and N. Isgur, Phys.Rev.D21:1868,1980  
\bibitem{d'angelo} A. d'Angelo, Talk at Baryons2002
\bibitem{buchmann}  A. Buchmann and E. Henley, Phys.Rev.D65:073017,2002
\bibitem{sato} T. Sato and T.S. Lee, T. Sato, this conferences
\bibitem{yang} S.S. Kamalov and S.N. Yang, Phys.Rev.Lett.83:4494-4497,1999 
\bibitem{beck} R. Beck et al., Phys.Rev.C61:035204,2000 
\bibitem{legs} G. Blanpied et al., Phys.Rev.C64:025203,2001 
\bibitem{frolov} V.V. Frolov et al., Phys.Rev.Lett.82:45-48,1999 
\bibitem{carlson} G. A. Warren, C.E. Carlson, Phys.Rev.D42:3020-3024,1990 
\bibitem{silva}A. Silva et al., Nucl.Phys.A675:637-657,2000 
\bibitem{warns} M. Warns, H. Schroder, W. Pfeil, H. Rollnik, 
Z.Phys.C45:627,1990 
\bibitem{liclose} Z.P. Li, F.E. Close,  Phys.Rev.D42:2194-2206,1990 
\bibitem{aznaury} I.G. Aznaurian,  Z.Phys.A346:297-305,1993 
\bibitem{leinweber} D. Leinweber, T. Draper, R.M. Woloshyn, Phys.Rev.D48:2230-2249,1993 
\bibitem{kjoo} K. Joo, et al, Phys. Rev. Lett. 88, 122001,2002
\bibitem{libuli} Z.P. Li, V. Burkert, Zh. Li; Phys.Rev.D46, 70, 1992
\bibitem{cano} F. Cano and P. Gonzales, Phys.Lett.B431:270-276,1998  
\bibitem{krewald} O. Krehl, C. Hanhart, S. Krewald, J. Speth, Phys.Rev.C62:025207,2000 
\bibitem{hovanes} H. Egiyan, Talk presented at Baryons2002 
\bibitem{lattice-1} S. Sasaki, T. Blum, S. Ohta  Phys.Rev.D65:074503 
\bibitem{lattice-2} D. Richards, private communications (2002)
\bibitem{lattice-3} W. Melnitchouk et al., hep-lat/0202022 
\bibitem{thompson} R. Thompson et al., Phys.Rev.Lett.86, 1702 (2001); 
H. Denizli, talk at Baryons2002
\bibitem{armstrong} C.S. Armstrong et al., Phys.Rev.D60:052004,1999 
\bibitem{janr} I. Aznaurian, private communications (2002)
\bibitem{saphir} M.Q. Tran et al., Phys.Lett.B445:20-26,1998 
\bibitem{santopinto} M.M. Giannini, E. Santopinto, A. Vassallo, Nucl.Phys.A699:308-311,2002; 
E. Santopinto, private communications (2002)
\bibitem{weise} N. Kaiser, P.B. Siegel, W. Weise, Phys.Lett.B362:23-28,1995 
\bibitem{hey} W.N. Cottingham and I.H. Dunbar, Z.Phys.C2, 41, 1979
\bibitem{ripani} M. Ripani, Nucl.Phys.A699:270-277,2002
\bibitem{capstick2} S. Capstick and W. Roberts, Phys.Rev.D49:4570-4586,1994 
\bibitem{kirchbach} M. Kirchbach, Mod. Phys. Lett. A12, 3177, 1997
\bibitem{mokeev} V.I. Mokeev, et al., Phys.Atom.Nucl.64:1292-1298,2001
\bibitem{page} S. Capstick, P.R. Page, Phys.Rev.D60:111501,1999  
\bibitem{pdg} D.E. Groom et al., Eur.Phys.J. C15, 1-878, 2000
\bibitem{cko} L.A. Copley, G. Karl, E. Obryk,  Nucl.Phys.B13:303-319,1969 
\bibitem{niculescu} G. Niculescu,  R. Feuerbach, private communications. 
\bibitem{asu} M. Duggar, B. Ritchie, et al., CLAS collaboration (PRL submitted)
\bibitem{fonvieille} H. Fonvieille, Talk given at Baryons2002
\bibitem{stepanya} S. Stepanyan et al., Phys.Rev.Lett.87,182002(2001)
\bibitem{hermes} A. Airapetian et al., Phys.Rev.Lett.87, 182001(2001)  
\bibitem{guidal} M. Guidal, Workshop on Electron Nucleus Scattering VII, 
Elba, Italy, June 24 - 27, 2002.

\end{thebibliography}
\end{document}